\documentstyle[eqsecnum,aps,epsfig]{revtex}
\newcommand{\tr}{{\rm tr}}
\newcommand{\bx}{{\bf x}}
\newcommand{\by}{{\bf y}}

\newcommand{\ri}{{\rm i}}
\newcommand{\re}{{\rm e}}

\newcommand{\sign}{{\rm sign}}
\newcommand{\mod}{{\rm mod}}
\begin{document}
\draft

\title{Spectral properties of Dissipative Chaotic Quantum Maps}
\author{Daniel Braun}
\address{FB7, Universit\"at--GHS Essen, 45\,117 Essen, Germany\\}

\maketitle
\thispagestyle{empty}
I examine spectral properties of a dissipative chaotic quantum map with the
help of a recently discovered semiclassical trace formula.  
I show that in the presence of a small amount of dissipation the traces of
any finite 
power of the propagator of the reduced density matrix, and traces of its
classical counterpart, the Frobenius--Perron operator, are {\em identical}
in the limit of 
$\hbar\to 0$. Numerically I find that even for finite $\hbar$ 
the agreement can be  very good. This holds in
particular if the classical phase space contains a
strange attractor, as long as one stays clear of bifurcations. Traces of
the quantum propagator for iterations of the map agree well with the
corresponding traces of the Frobenius--Perron operator if the classical
dynamics is dominated by a strong point attractor.  
\bigskip

\section{Introduction}\label{intro}
The interplay between  chaos, quantum mechanics and  dissipation is 
rather complex and the subject of strong current research activities
\cite{Dittrich98,Habib98,Cohen97,Dittrich90,Graham86,Graham85}. 
The definition of chaos in classical
mechanics via exponentially fast spreading trajectories can not be applied
to quantum mechanical systems, since the notion 
of a trajectory does not exist in quantum mechanics. On a quantum mechanical
level chaos manifests itself in the statistical properties of the
eigenenergies and eigenfunctions. In the case of Hamiltonian
systems the eigenenergies and eigenfunctions obey
the universal statistics of large random hermitian matrices restricted only
by general 
symmetry requirements like invariance under time and spin reversal
\cite{Bohigas84,Berry83}. While no rigorous proof of this conjecture is
known yet, overwhelming numerical and experimental evidence has been
accumulated \cite{Guhr98,Reichl92,Haake91}.\\  
Dissipation has at least two very important effects. The classical dynamics
is altered profoundly. It is no longer 
restricted to a shell of constant energy in phase space, but phase space
volume shrinks if no external source compensates for the energy
dissipated. In a chaotic system with external driving dissipation typically
leads to a 
strange attractor in phase space, i.e.~a multi-fractal structure that is
invariant under 
the dynamics and which has a dimension strictly smaller than the dimension
of the 
phase space.  The second effect is of quantum mechanical nature and 
more subtle: Dissipation destroys very
efficiently the
quantum mechanical phase information. This typically happens on time scales
much shorter than classical ones and even with very tiny amounts of
dissipation \cite{Zurek81,Giulini96,DBraun98.2}. Therefore the system
behaves more 
classically, and one might expect to find classical
manifestations of chaos again. It was indeed shown in a variety of examples
that appropriate quantum mechanical counterparts of the classical phase
space density (like Husimi functions or Wigner distributions) approach a
smeared out version of the strange attractor
\cite{Dittrich90,Peplowski91,Kolovsky96}. At the same time one might ask
whether spectral properties approach their classical
counterparts as well. 
This paper shows that for certain spectral properties the answer is
``YES'', even though the structure of the
spectrum can be very different in both cases. \\  

My
analysis is based on a recently discovered trace formula for
dissipative systems which, in the spirit of  Gutzwiller's
celebrated formula \cite{Gutz70,Gutz71}, expresses traces of the
propagator of the density matrix in terms of classical periodic orbits
\cite{DBraun98}. I  show in section \ref{sectrace} that to lowest order 
asymptotic expansion in $\hbar$ the traces agree with the traces of
the classical Frobenius--Perron propagator of the phase space density
\cite{Cvitanovic91}.\\ 
In
the next section I briefly review basic properties of dissipative quantum
maps, semiclassical theory and the trace formula. In section
\ref{secnum} I apply the trace formula to a dissipative kicked top and
compare with numerical results for finite $\hbar$. The main
results are summarized in section \ref{summary}.

\section{Dissipative Quantum Maps}
\subsection{General remarks}
Dissipation is introduced on a quantum mechanical level most rigorously by
the so--called Hamiltonian embedding \cite{Weiss93}. The system of interest
is considered 
as part of a larger system including the ``environment'' to which
energy can be dissipated. The total system is assumed to be
closed so that it is adequately described by a Schr\"odinger equation. The
degrees of freedom of the ``environment'' remain unobserved. The
system at interest is described by a density matrix in which the
environmental degrees of  
freedom have been traced out, usually termed the reduced density matrix
$\rho(t)$. \\
Dissipative quantum maps are maps of the reduced density matrix from a time
$t$ to a time $t+T$: $\rho(t+T)=P\rho(t)$. They are analogous to
non--dissipative quantum maps, where the state vector of the system is
mapped with a unitary Floquet matrix $F$, $|\psi(t+T)\rangle
=F|\psi(t)\rangle$. In the dissipative case $P$ is not a unitary operator
and therefore has eigenvalues inside the unit circle.\\
Maps, dissipative or not, are a natural 
description of a time evolution if an external driving of the system is
periodic in time with period $T$. They give a stroboscopic picture which
suffices if the evolution during one period is of no 
interest. Systems that are
periodically driven are 
capable of chaos even if they have only one degree of freedom. I will
restrict myself in the following to such cases. 

The maps that I consider are particularly simple
in the sense that the dissipation is well separated from a
remaining purely unitary evolution where the latter by itself is capable of
chaos. The unitary part will be 
 described by a Floquet matrix $F$ acting on the state vector, so that the
unitary evolution takes the density matrix from $\rho(t)$ to
$\rho'(t)=F\rho(t)F^\dagger$. After the unitary part a
dissipative step follows which I will describe by  a  propagator $D$. It
takes the density matrix from $\rho'(t)$ to $\rho(t+T)=D\rho'(t)$.
The total map therefore reads
\begin{equation} \label{map}
\rho(t+T)=D(F\rho(t) F^\dagger)\equiv P\rho(t)\,.
\end{equation} 
Such a separation into two parts is not purely academic. A most obvious
realization of (\ref{map}) is given when a Hamiltonian $H(t)$ leading to the
unitary evolution and the coupling to the environment can be turned on and
off alternatively. This should be realizable for instance with atoms flying
through a series of cavities where in each cavity either the unitary
evolution or the dissipation is realized. Another example might be a
billiard, in which the particle only dissipates energy
when hitting the 
walls. But even if the dissipation cannot be turned off, the map
(\ref{map}) may still be a good description. For instance, if
the dissipation is weak and if the entire unitary evolution takes place
during a very short  
time, dissipation may be negligible during that time. This is the case
if the entire unitary evolution is due to a periodic kicking. The dissipation
can then be considered as a relaxation process between two successive
kicking events.
 Finally, a formal reason for such a separation can be given when
the generators for the unitary evolution and the dissipation commute.

These ideas might become clearer with a particular model.  Let me
therefore introduce as prime model system a dissipative kicked top.\\

\subsection{A dissipative kicked top}
The dynamical variables of a top \cite{kickedtop,Haake91} 
are the three components $J_{x,y,z}$ of an angular momentum ${\bf J}$. I
will only consider dynamics (including the dissipative ones) which conserve
the absolute value of ${\bf 
J}$, ${\bf J}^2=j(j+1)=const$. In the classical limit 
(formally attained by letting the quantum number $j$ approach infinity) the
surface of the unit sphere $\lim_{j\rightarrow\infty} ({\bf J}/j)^2=1$
becomes the phase space, such that one confronts but a single degree of
freedom. Convenient phase space coordinates are 
\begin{equation} \label{pq}
\mu\equiv J_z/J=\cos\theta=p\mbox{ and } \varphi=q, 
\end{equation} 
where the polar and azimuthal angles $\theta$ and $\varphi$ 
define the orientation of the angular momentum vector with respect to the
$J_x$, $J_y$ and $J_z$ axes. The parameter $J$ is defined as $J=j+1/2$ and
allows for more convenient expressions of most semiclassical quantities. 
Due to the conservation of ${\bf J}^2$ the Hilbert
space decomposes into $(2j+1)$ dimensional subspaces. The
quantum dynamics is confined to one of these according to the initial
conditions.  
The semiclassical limit is characterized
by large values  of the quantum number $j$ which can be integer or half
integer.  Since the classical phase space
contains $(2j+1)$ states, Planck's constant may be thought of as
represented by $1/J$. \\

Consider a unitary evolution generated by the
Floquet matrix  
\begin{equation} \label{F}
F=\re^{-\ri \frac{k}{2J}J_z^2}\re^{-\ri \beta J_y}\,.
\end{equation}
The corresponding classical motion first rotates the  angular momentum by an
angle $\beta$ about the $y$-axis and then subjects it to a torsion about the
$z$-axis. The latter may be considered as a non--linear rotation  with a
rotation angle   
given by the $J_z$ component of ${\bf J}$. The dynamics is known to become
strongly chaotic for sufficiently large values of $k$ and $\beta$, whereas
either $k=0$ or $\beta=0$ lead to integrable motion \cite{Haake91}. For
a physical realization of this dynamics it might be best to think of
${\bf J}$ as a Bloch vector describing the collective excitations of
two--level atoms, as one is used to in quantum optics. The rotation can be
brought about by a laser pulse of suitably chosen length and intensity,
and the 
torsion by a cavity that is strongly off resonance from
the atomic transition frequency \cite{Agarwal97}. The Floquet matrix
(\ref{F}) has 
also been realized in experiments with magnetic crystallites with an easy
plane of magnetization \cite{Waldner85}.

Our model dissipation is defined in
continuous time $\tau$ by the Markovian master equation
\begin{equation}
\label{rhotd}
\frac{d}{d\tau}\rho(\tau)=\frac{1}{2J}([J_-,\rho(\tau)
J_+]+[J_-\rho(\tau), J_+])\equiv \Lambda\rho(\tau)\,,
\end{equation}
where the linear operator $\Lambda$ is defined by this equation as generator
of the dissipative motion. 
Equation (\ref{rhotd}) is well-known to describe certain superradiance
experiments, 
where a large number of  two--level atoms in a cavity of bad quality radiate
collectively 
\cite{Bonifacio71.1,Haroche82}. The angular momentum operator ${\bf J}$ is
then again the Bloch vector of the collective excitation and the $J_+,J_-$
are raising and lowering operators, $J_\pm=J_x\pm\ri J_y$.
One easily verifies that (\ref{rhotd})
conserves the skewness in the $J_z$ basis ($J_z|m\rangle =m|m\rangle$),
i.e.~matrix elements $\langle m_1 |\frac{d}{d\tau}\rho|m_2\rangle$ with
a given 
skewness $J\eta=m_1-m_2$ depend 
only on matrix elements with the same
skewness. Eq.(\ref{rhotd}) is formally solved by
$\rho(\tau)=\exp(\Lambda\tau)\rho(0)$ for any initial
density matrix $\rho(0)$ and this defines the 
dissipative propagator 
\begin{equation} \label{D}
D(\tau)=\exp(\Lambda\tau)\,.
\end{equation}
Explicit forms of $D$ can be found in
 \cite{Bonifacio71.1,PBraun98.1,PBraun98.2}.  
 The skewness $\eta$ only enters as a parameter in
$D$.  
The classical limit gives the simple picture of the Bloch vector creeping
towards the south pole $\theta=\pi$ as an over-damped
pendulum, according to the equations of motion
\begin{equation} \label{eomdiss}
\frac{d}{d\tau}\theta=\sin\theta\mbox{, }\frac{d}{d\tau}\varphi=0\,.
\end{equation}
Classically the azimuth $\varphi$ is therefore
conserved. Eq.(\ref{eomdiss}) also shows that $\tau$ is the time in units of
the classical time scale. In the following it will be set equal to the time
between two unitary steps. \\

The Floquet matrix (\ref{F}) is usually generated by the Hamiltonian
\begin{equation} \label{Hoft}
H(t)=\hbar\left(\frac{k}{2JT}J_z^2+\beta J_y\sum_{n=-\infty}^\infty\delta(t-nT)\right);
\end{equation}
it describes the evolution from immediately before a kick to immediately
before the next kick. The generator (\ref{rhotd}) for the dissipation does
not commute with $H(t)$. In order to obtain the map (\ref{map}) one should
replace in (\ref{Hoft}) $H_0=\frac{k}{2JT}J_z^2$ by $\frac{k}{2JT_1}J_z^2$
and switch on $H(t)$ only for a time $T_1<T$ during each period $T$, whereas
the dissipation acts during the rest of the time
$\tau=T-T_1$. Alternatively, when $H(t)$ and $\Lambda$ act permanently one
may go to an interaction representation by
$\rho(t)=
\exp(-\frac{\ri}{\hbar}H_0t)\tilde{\rho}(t)\exp(\frac{\ri}{\hbar}H_0t)$. In
the $J_z$ representation this leads only to phase factors in the master
equation (\ref{rhotd}) which can be easily incorporated in $D$ and which
vanish moreover for diagonal elements. Let us assume in the following that
either has been done and use (\ref{map}) with $F$ and $D$ given by
(\ref{F}), (\ref{rhotd}), and (\ref{D}) as a starting point with $\tau$ as
fixed parameter that measures the relaxation time between two unitary
evolution and thus the dissipation strength.

\subsection{The trace formula}
In 1970 Gutzwiller published a trace formula for Hamiltonian flows
that has become a center piece of subsequent studies of quantum chaotic
systems \cite{Gutz70,Gutz71}. A corresponding formula was obtained later for
non-dissipative quantum maps by Tabor \cite{Tabor83}. Assuming the existence
of a corresponding classical map ${\bf y}={\bf
f}({\bf x})$ of phase on itself ($\bx=(p',q')$ are the old, $\by=(p,q)$ the
new phase space coordinates), both
formulae express a spectral property of the quantum mechanical propagator
as a sum over periodic 
orbits of ${\bf f}$. Each periodic orbit
contributes a weighted phase factor, where the weight depends on the
stability matrix $M$ of the orbit and the phase is basically given by the
classical 
action $S$ in units of $\hbar$. Tabor's formula aims at 
traces of the Floquet matrix $F$,
\begin{equation} \label{trFN}
\tr F^N=\sum_{p.p.}\frac{\re^{\ri(\frac{
S}{\hbar}-\frac{\pi}{2}\nu)}}{|2-\tr M|^{1/2}}\,.
\end{equation}
I have written the sum over periodic orbits as a sum over periodic
points ($p.p.$) of the $N$ times iterated map ${\bf
f}^N$; the integer $\nu$ (the
so--called Maslov index)
counts the number of caustics along the orbit.
 All quantities have to be evaluated on the periodic points.
The squared modulus of $\tr F^N$ has in the unitary case an interpretation
as (discrete time) form factor of spectral correlations.\\
 
In  \cite{DBraun98} a corresponding 
trace formula for dissipative quantum maps of the form (\ref{map})  was
derived. 
It is based on semiclassical approximations for both $F$
and $D$. The semiclassical approximation of $F$ 
has the general form of a  
van Vleck propagator \cite{PBraun96,vanVleck28}; a corresponding
semiclassical approximation for $D$ was obtained in \cite{PBraun98.2}. A
WKB ansatz 
lead to a fictitious Hamiltonian
system which depends on the skewness as a parameter. Its trajectories 
connect initial and final points specified by the 
arguments of $D$. Much as in the unitary case, an action $R$  is 
accumulated along the trajectories; it has the usual
generating properties of an action. Based only on the general van Vleck
forms of 
$F$ and $D$ and the generating properties of the actions $S$ and $R$ 
we derived the  trace formula
\begin{equation} \label{trPN}
\tr P^N=\sum_{p.p.}\frac{\re^{\sum_{i=1}^N JR_i}}{\left|\tr
\prod_{i=N}^1M_d^{(i)}-\tr 
M\right|}\mbox{, }\quad\quad N=1,2,\ldots\,.
\end{equation} 
The sum is over all
periodic points of the $N$--times iterated dissipative classical map; the $R_i$
are the actions of the fictitious Hamiltonian system for vanishing skewness
accumulated during  the
$i$th dissipative step.
 The denominator contains the stability matrices 
$M_d^{(i)}$ for the $i$th dissipative step and $M$ for the
entire map ${\bf f}^N$. The matrix $M_d^{(i)}$ with index $i=N$ is at the
left of the product. 
Eq.(\ref{trPN}) is a leading order asymptotic expansion in $1/J$ for
propagators $P$ of the type (\ref{map}). The 
following restrictions apply:
\begin{itemize}
\item The phase space is two dimensional.
\item The classical limit of the dissipative part of the map conserves one
phase space coordinate (the azimuthal coordinate  $\varphi$ for the
dissipation described by (\ref{rhotd})). 
\item The propagator $D$ for the dissipative part conserves the
skewness $\eta$ of the density matrix in a suitably chosen basis and $D$ has
a single maximum as a function of $\eta$ at $\eta=0$. As indicated the
dissipation (\ref{rhotd}) conserves the skewness in the $J_z$ basis.
\item The dissipation exceeds a certain minimum value. It is given by
$\tau\gtrsim 1/J$ for the dissipation (\ref{rhotd}) and thus may become
infinitesimally small in the classical limit $J\to\infty$.
\end{itemize}
   
Eq.(\ref{trPN}) shows that periodic orbits and classical
quantities related to them still determine  the  spectral properties of
the quantum system even in the presence of dissipation.  The formula will
now be studied in more detail.

\section{Connection to classical trace formula}\label{sectrace}
Remarkable about
(\ref{trPN}) is its  simplicity. First of all, when propagating a density
matrix, one would expect a 
double sum over periodic points. Indeed, in the dissipation free case one
 easily shows $\tr P=|\tr F|^2$, and $\tr F$ is given by the Tabor formula
(\ref{trFN}) 
 as a simple sum over
periodic points \cite{Tabor83}. Out of the double sum, only the ``diagonal
parts'' survive. Decoherence induced  through dissipation  destroys the
interference terms between different periodic points. For the diagonal terms
the actions $S$ and $-S$ stemming from $F$ 
and $F^\dagger$ and the
phases due to the Morse indices cancel. The square
roots in the denominator combine to a power 1. Due to the cancellation of
the phase factors the traces
(\ref{trPN}) are always real and
positive. They fulfill herewith a general requirement for all propagators of
density matrices that follows from conservation
of positivity of the density matrix. 
On the other hand one may wonder whether the trace formula should not be an
entirely classical formula, if all interference terms are
destroyed. This is indeed what I am going to show now.

The classical propagator of phase space density is given by
$P_{cl}(\by,\bx)=\delta(\by-{\bf f}({\bf x}))$. In the case where
$P_{cl}(\by,\bx)$ describes the map arising from the evolution during a
finite time of an autonomous system, $P_{cl}$ is  commonly called the
Frobenius--Perron operator. For brevity I use the same name in the present
dissipative situation. The trace of the $N$th iteration of $P_{cl}$ is
given by \cite{Cvitanovic91}
\begin{eqnarray}
\tr P_{cl}^N&=&\sum_{p.o.}\sum_{r=1}^\infty
\frac{n_p\delta_{N,n_pr}}{|\det({\bf 
1}-M_p^r)|} \label{first}\\
&=&\sum_{p.p.}\frac{1}{|\det({\bf 1}-M)|}\,,\label{second}
\end{eqnarray}   
where the first sum in (\ref{first}) is over all primitive periodic orbits of
length 
$n_p$, $r$ is their repetition number and $M_p$ the stability matrix of the
primitive orbit. In (\ref{second}), $p.p.$ labels all periodic points
belonging to 
a periodic orbit of total length $N$, including the repetitions, and $M$ is
the stability matrix for the entire orbit. \\
The fact that $M$ in (\ref{trPN}) is a $2\times 2$ matrix leads immediately to
$\det({\bf 1}-M)=1+\det M-\tr M$. Since the map is a periodic succession of
unitary evolutions (with stability matrices $M_u^{(i)}$)  and dissipative
evolutions (with stability matrices $M_d^{(i)}$), $M$ is given by the
product $M=\prod_{i=N}^1M_d^{(i)}M_u^{(i)}$. The stability matrices
$M_u^{(i)}$ are all unitary so that $\det M_u^{(i)}=1$ for all $i=1\ldots
N$ and $\det
M=\prod_{i=N}^1\det M_d^{(i)}$. The dissipative process for which
(\ref{trPN}) was derived 
conserves $q$ which means that $M_d^{(i)}$ is diagonal,
\begin{equation} \label{md}
M_d^{(i)}=\left(\begin{array}{cc}1 &0\\
	0&m_d^{(i)}
\end{array}
\right)\,.
\end{equation}
The upper left element is $\frac{\partial q(p',q')}{\partial q'}$, the lower
right $\frac{\partial p(p',q')}{\partial p'}$.
But then $\det M_d^{(i)}=m_d^{(i)}$, and we find with $\left|\tr \prod_{i=N}^1 M_d^{(i)}-\tr M\right|=|1+\det 
M-\tr M|= |\det({\bf
1}-M)|$ exactly the
denominator 
in (\ref{second}). \\
The actions $R_i$  are zero on the
classical trajectories for the dissipative process (\ref{rhotd}), as one
immediately sees by using their explicit form
\cite{PBraun98.2}. Their vanishing can be retraced more
generally to conservation of probability by the master equation and
therefore holds for other master equations of the same structure as
well. To see this write (\ref{rhotd}) in the $J_z$ basis and look at the
part with vanishing skewness, i.e.~the 
probabilities  $p_m=\langle
m|\rho|m\rangle$. We obtain a set
of equations
\begin{equation} \label{pt}
\frac{d}{d\tau}p_m=(g_{m+1}p_{m+1}-g_m p_m)\,,
\end{equation} 
where the specific form of the coefficients $g_m$ is of no further
concern. Important is rather that the {\em same} function $g_m$ appears
twice. This is sufficient and necessary for the conservation of
probability, $\tr
\rho=\sum_{m=-j}^jp_m=1$. On the other hand, had we coefficients $f_m$ and
$g_m$ 
(i.e.~$\dot{p}_m=(g_{m+1}p_{m+1}-f_m p_m)$) we would obtain the action
$R$  on the classical trajectory as $JR=\sum_{l=m}^n(\ln(g_l)-\ln(f_l))$ as
one easily verifies by writing down the exact Laplace image of $D$ following
the lines in \cite{PBraun98.1}. Thus,
the action is zero iff probability is conserved.
But then {\em the trace formula (\ref{trPN}) is 
identical to the classical trace formula (\ref{second}).}

This result proves that the
traces of any finite power of the evolution operator of the quantum
mechanical density matrix, are, in the limit of $\hbar\to 0$, exactly given
by the corresponding traces of the evolution operator of the classical phase
space density, provided a small
amount of dissipation is introduced. This is quite surprising since it is
clear that even the basic structure of the two spectra can be very
different: For all finite Hilbert
space dimensions $d=2j+1$ the quantum mechanical propagator $P$ can be
represented as a finite $d^2\times d^2$ matrix. Its 
spectrum is therefore always discrete, regardless of whether the
corresponding classical map is chaotic or not. On the other hand, it is
known that $P_{cl}$ has necessarily a continuous spectrum if the classical
dynamics is mixing \cite{Gaspardbook}.

A formal reason why the spectra may differ in
spite of the fact that the traces agree to lowest order in $\hbar$ is easily
found. In order to 
construct the entire spectrum of $P$ one needs $d^2=(2j+1)^2$ traces. But
already for traces of 
order $j$ the next order corrections
in the asymptotic expansion in $1/J$ that lead to (\ref{trPN}) become
comparable to the classical term; and for the highest
traces needed (i.e. traces of order $j^2$) the next order in $1/J$ would
be even more important than the classical term, so that one may not expect $\tr
P^{j^2}=\tr P_{cl}^{j^2}$ for $j\to\infty$. In other words, if we do {\em not}
keep $n$ fixed in the classical limit, $ \tr
P^n=\tr P_{cl}^n$ may not hold for $j\to \infty$ and therefore the spectrum
of $P$ can be 
very different from that of $P_{cl}$. \\  

The asymptotic equality of $\tr P^n$ and $\tr P_{cl}^n$ strengthens
substantially the quite envolved semiclassical derivation of 
(\ref{trPN}) \cite{DBraun98}. It also sheds light
on the question what happens if the dissipation does not conserve the
coordinate $q$. We should then not expect (\ref{trPN}) to be valid 
but presumably replace it with the more general form  (\ref{second}).   

\section{Comparison with numerical results}\label{secnum}

The
question arises how good the agreement between quantum and classical  traces
is for finite $J$. To answer this 
question I have calculated numerically the exact quantum mechanical traces for
our dissipative kicked top and compared them with the traces
obtained from  the trace formula (\ref{trPN}). These
results will be presented now.

\subsection{The first trace}
\begin{figure}
\epsfig{file=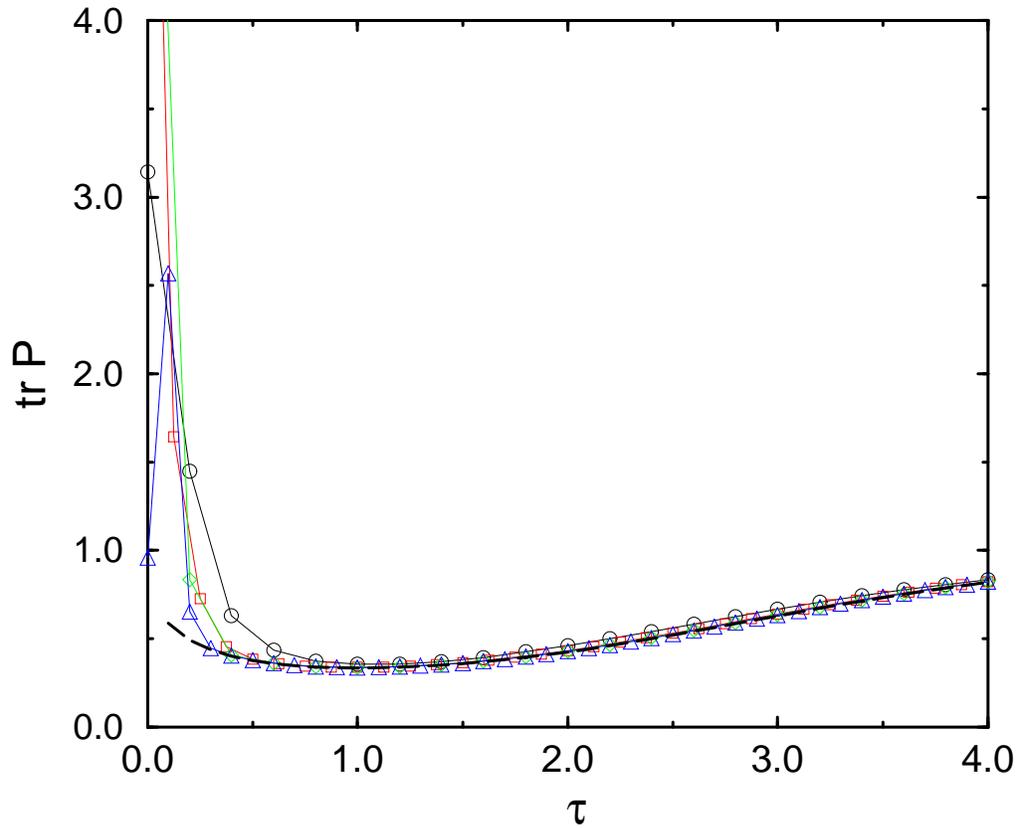,width=15cm}
\protect\caption{\label{figtrPk4b2} Comparison of quantum mechanical traces
($j=10$ (circles), $j=30$ (squares), $j=50$ (diamonds) and $j=80$
(triangles)) with 
classical trace (thick dashed line) for $k=4.0$, $\beta=2.0$ as function of
$\tau$. There are no bifurcations for $\tau\protect\gtrsim 0.1$.}   
\end{figure}

The quantum mechanical propagator $P$ is most conveniently calculated in the
$J_z$ basis, since the torsion part is then already diagonal. The rotation
about the $y$--axis leads to a Wigner $d$--function whose values
are obtained numerically via a recursion relation as described in
\cite{PBraun96}. The propagator for the dissipation is obtained 
by inverting numerically the exactly known Laplace image
\cite{Bonifacio71.1,PBraun98.1}. The total
propagator $P$ is a full, complex, non--hermitian, and non--unitary matrix of
dimension $(2j+1)^2\times (2j+1)^2$. Since for the first trace the knowledge
of the diagonal matrix elements suffices I was able to calculate
$\tr P$ up to $j=80$. Higher traces are most efficiently obtained via
diagonalization which limited the numerics to $j\le 40$.\\
The effort for calculating the first classical trace is comparatively
small. In all examples considered and even in the presence of a strange
attractor, $P_{cl}$ had at most 4 
fixed points that could easily be found
numerically by a simple Newton--method in two dimensions. For each fixed
point the 
stability matrix is found via the formulae in Appendix \ref{appA} and so
the trace is immediately obtained .\\
In Fig.\ref{figtrPk4b2} I show $\tr P$ for different values of $j$ as a
function of $\tau$ and compare with $\tr P_{cl}$, eq.(\ref{trPN}). The
values for torsion strength and rotation angle, $k=4.0$ and $\beta=2.0$
were chosen such that the system is already rather chaotic in the dissipation
free 
case at $\tau=0$; a phase space portrait of many iterations of $P_{cl}$ shows
a large chaotic sea and 6 relatively small stable islands. When $\tau$
reaches a value of the order $\tau\simeq 
0.5$ a strange attractor appears  which
rapidly changes its form and 
dimension when $\tau$ is increased. The attractor shrinks and is pushed more
and more towards the south pole, as the  angular
momentum has more and more time to relax towards the 
ground state $J_z=-j$ between two kicks.
At values of $\tau$ of the order of $\tau\simeq 2.0-3.0$ the
attractor degenerates to a strong point attractor close to the south pole
which absorbs even 
remote initial points in very few steps. At even stronger damping the
point attractor reaches the south pole asymptotically.\\
Figure \ref{figtrPk4b2} shows that -- with the exception of very small
damping -- $\tr 
P_{cl}$ reproduces $\tr P$ perfectly well for all $\tau$, in
spite of the strongly changing phase space structure. The agreement extends
to smaller $\tau$ with increasing $j$, as is to be expected from
the condition of validity of the semiclassical approximation,  $\tau\gtrsim
1/J$ 
\cite{PBraun98.2}. The analysis of the fixed points shows that at $k=4.0$,
$\beta=2.0$ always two fixed points exist for $\tau\gtrsim 0.1$. Their $\mu$
component slowly decreases and the lower one converges towards the south pole
with increasing $\tau$, where it finally coincides with the point attractor. \\

\begin{figure}
\epsfig{file=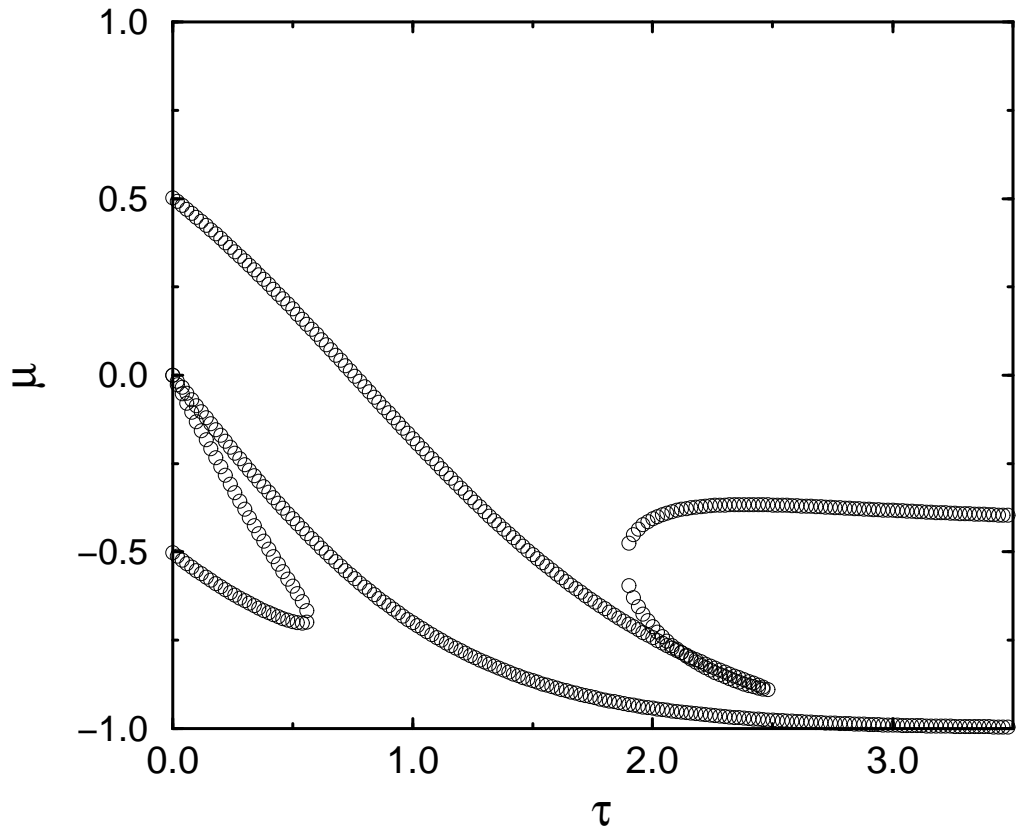,width=15cm}
\protect\caption{\label{figfixedpk8b2} The $\mu$--component of the fixed
points at $k=8.0$, $\beta=2.0$ as a function of  $\tau$. There are four fixed points at $\tau=0.0$ out of which two coincide and
disappear at $\tau\simeq 0.57$. A new pair is born at
$\tau\simeq 1.89$, but one fixed point disappears again at
$\tau\simeq 2.47$, in close vicinity with one of the original fixed points.}  
\end{figure}

\begin{figure}
\epsfig{file=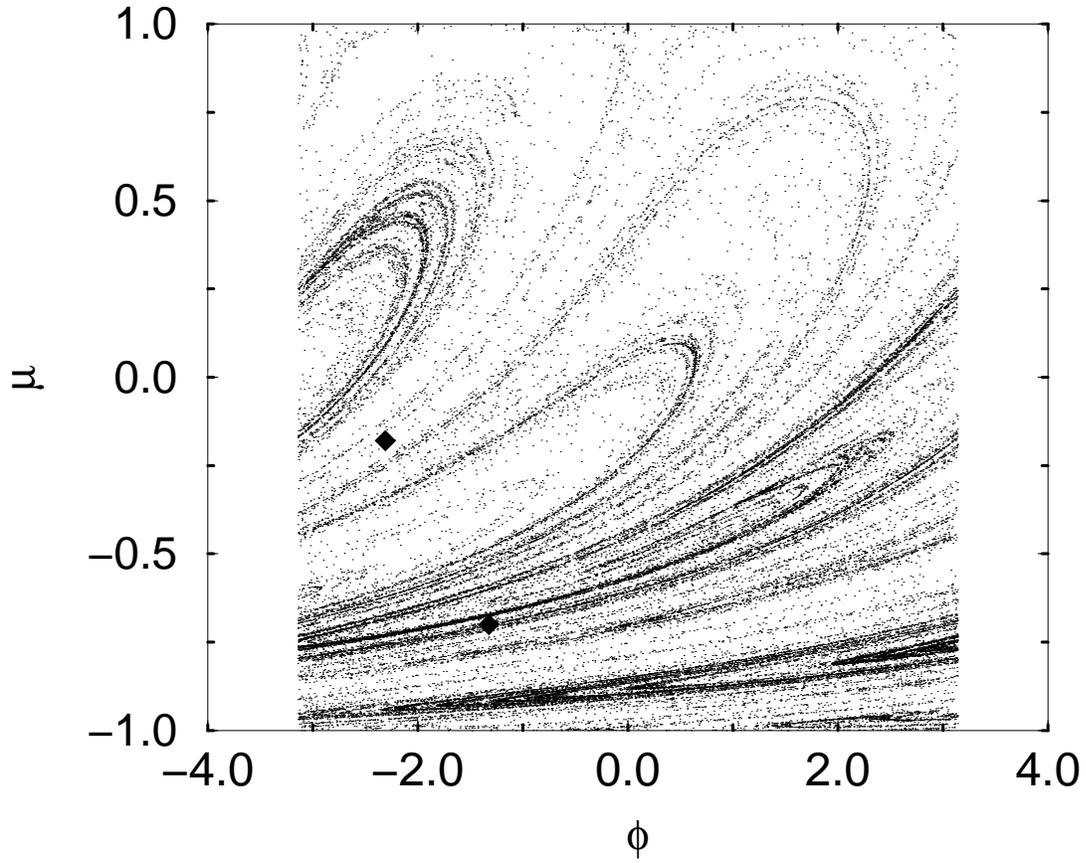,width=15cm}
\protect\caption{\label{figSAk8b2t1} Strange attractor for $k=8.0$,
$\beta=2.0$, $\tau=1.0$. The diamonds mark the position of the two fixed
points. The borders $\varphi=\pi$ and $\varphi=-\pi$ have to be identified.}  
\end{figure}

Fig.\ref{figfixedpk8b2} shows the fixed point structure for a more
complicated situation ($k=8.0$, $\beta=2.0$). The dissipation free dynamics
at $\tau=0$ is entirely chaotic, no visible phase space structure is
left. The 
above statements about the creation of a strange attractor (see
Fig.\ref{figSAk8b2t1}) and its
degeneration to a point attractor when $\tau$ is increased apply equally well. 
 
\begin{figure}
\epsfig{file=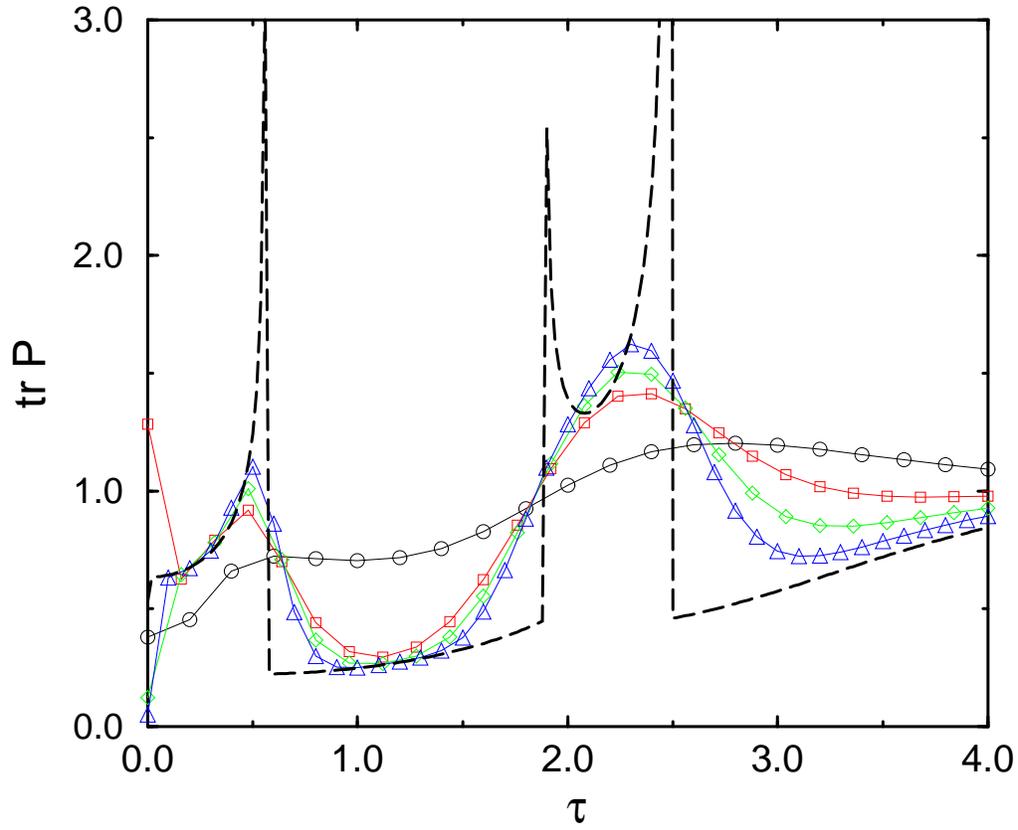,width=15cm}
\protect\caption{\label{figtrPk8b2} Comparison of quantum mechanical traces
 with
classical trace for  $k=8.0$,
$\beta=2.0$ as a function of 
$\tau$ (same symbols as in Fig.\ref{figtrPk4b2}). The classical trace
diverges whenever a bifurcation is reached.}   
\end{figure}

In Fig.\ref{figtrPk8b2} I show the first trace as function of $\tau$ for 
this situation. The classical trace diverges whenever a bifurcation is
reached. Such a behavior is well known from the Gutzwiller formula in the
unitary case; 
the reason for the divergence is easily identified as breakdown of the
saddle point approximation in the semiclassical
derivation of the trace formula. Whereas the 
quantum mechanical traces for small $j$ (say $j\simeq 10$) seem not to take
notice of the bifurcations, they approximate the jumps and divergences
better and better when $j$ is increased. At $j=80$ the agreement with the
classical trace is already very
good between the
bifurcations. Remarkable, however, 
is the fact that there are some values of $\tau$ close to the
bifurcations, where all $\tr P$ curves for different $j$ in the entire $j$
range examined cross. The trace
seems to be independent of $j$ at these points, but
they nevertheless do not lie on the classical curve. One is reminded of a
Gibbs phenomenon, but I do not have any explanation for it.  

\subsection{Higher Traces}
\begin{figure}
\epsfig{file=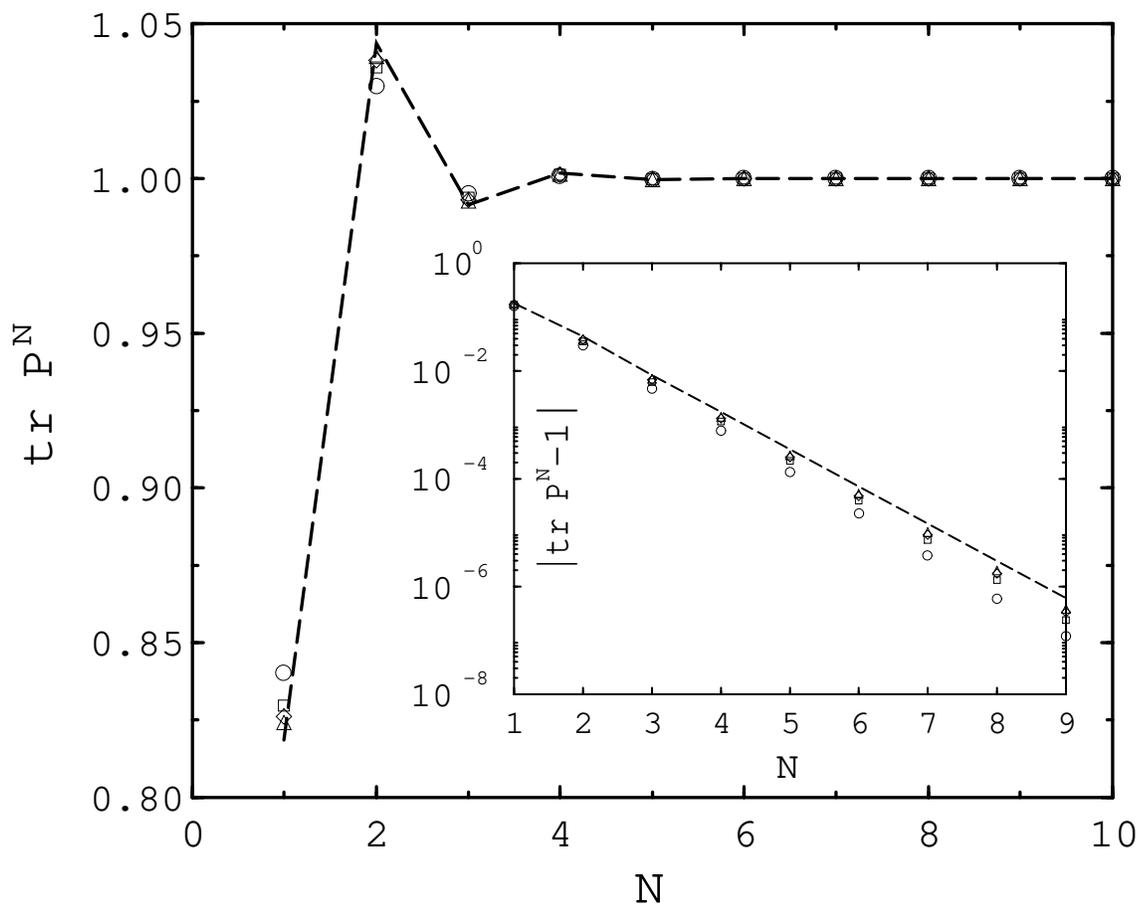,width=15cm}
\protect\caption{\label{figtrPNk4b2t4} Quantum mechanical  and classical
trace as a function of $N$ for 
$k=4.0$, 
$\beta=2.0$, $\tau=4.0$ (same symbols as in Fig.\ref{figtrPk4b2}). The
classical trace is shown as dashed line for 
better visibility, even though it is only defined for integer $N$. The inset
shows that the exponential convergence to 
1 also holds in the classical case. The classical dynamics is dominated 
by a single point--attractor/repeller pair.} 
\end{figure}

\begin{figure}
\epsfig{file=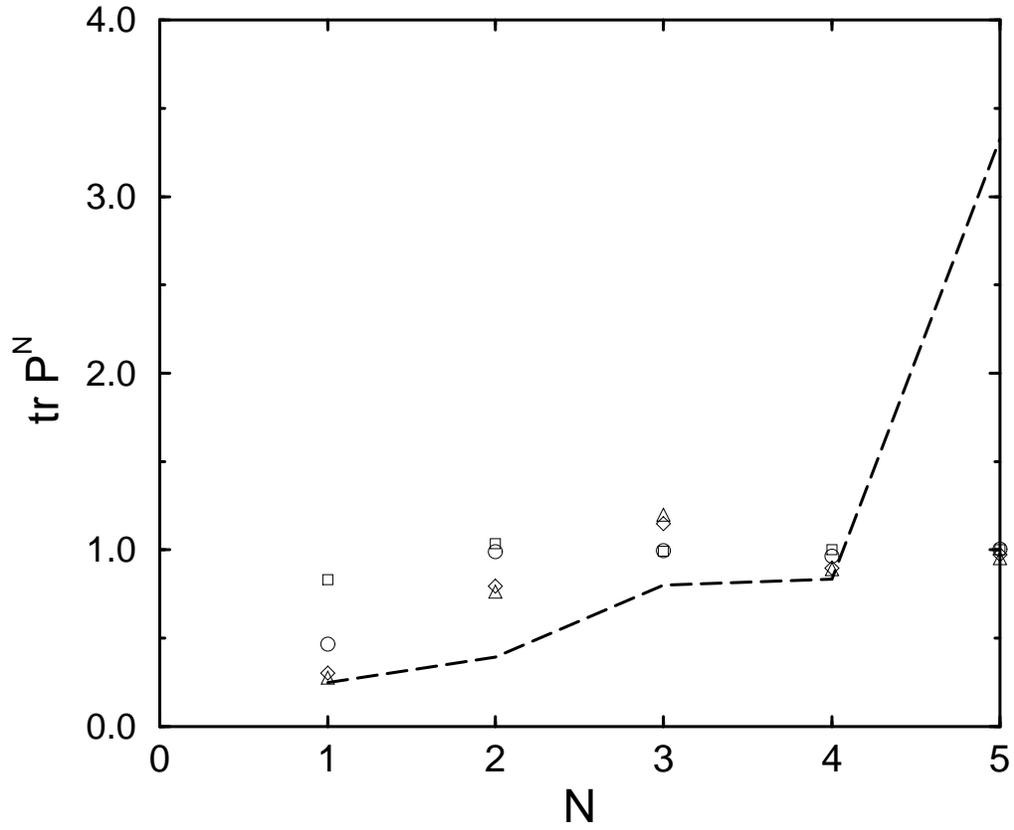,width=15cm}
\protect\caption{\label{figtrPNk8b2t1} Quantum mechanical  and classical
trace as a function of $N$ for $k=8.0$, 
$\beta=2.0$, $\tau=1.0$ (same symbols as in Fig.\ref{figtrPk4b2}). The
corresponding phase space portrait is the strange 
attractor shown in Fig\ref{figSAk8b2t1}.}
\end{figure}

Let us now examine higher traces $\tr P^N$ for given values of $k$, $\beta$,
and 
$\tau$ as a function of $N$. For large $N$ all higher traces must converge
exponentially to $1$, independent of the 
system parameters. This is due to the fact that $P$ has always one
eigenvalue equal to $1$. Its existence follows from elementary
probability conservation \cite{Haake91}. The corresponding eigenmode is an
invariant density matrix, its 
classical counterpart the (strange or point) attractor, the fixed points or
any linear combinations thereof \cite{Gaspardbook}. 
All other eigenvalues have an absolute value smaller than $1$
since there is only dissipation and no amplification in the system. Their
powers decay to zero as a function of $N$.\\ 
I will focus on two limiting cases: The case where the basic phase space
structure is a point attractor and the case where it is a well
extended strange attractor. 
As explained above a point attractor can always be obtained by sufficiently
strong damping. Consider the example $k=4.0$, $\beta=2.0$ and $\tau=4.0$. 
Fig.\ref{figtrPNk4b2t4} shows that indeed both quantum mechanical and
classical result converge rapidly towards $1$, and the agreement is very
good even for $j=10$. If one examines the convergence rate one
finds that it is slightly $j$-dependent, but rapidly reaches the classical
value. It should be noted that the calculation of $\tr P_{cl}^N$ is
enormously simplified here by the fact that with increasing $N$ no
additional periodic points arise. The dissipation is so strong that the
system is integrable again. In the example given there are only
two fixed points, one at $(\mu,\phi)\simeq  (-0.3812219, -3.098751)$, a strong
point repeller, and one at $(\mu,\phi)\simeq  (-0.9984018,-1.444154)$ a strong
point attractor, and all periodic points of 
$P_{cl}^N$ are just repetitions of these two points. \\
The situation is quite different in the case of a strange attractor
(Fig.\ref{figtrPNk8b2t1}). The 
number of periodic points increases exponentially 
with $N$, as is typical for chaotic systems. This makes the classical
calculation 
of higher traces exceedingly difficult. For $k=8.0$, $\beta=2.0$,
and $\tau=1.0$ I was able to calculate $\tr P_{cl}^N$ reliably up to
$N=5$, where about 400 periodic points have to be taken into
account. The obtained numerical result for $\tr P_{cl}^N$ can
always be considered as lower bound for the exact result for $\tr P_{cl}^N$
as long as one can exclude over-counting of fixed points since all terms in
the sum (\ref{trPN}) are positive. It is then clear that at $N=5$ the
quantum mechanical 
result at $j=40$ is still more than a factor 3 away from $\tr
P_{cl}^N$, even though for $N=1$ the agreement is very good. The
convergence of $\tr P^N$ to $\tr P_{cl}^N$ as a function of $j$ becomes
obviously worse with increasing $N$. \\

\section{Summary}\label{summary}

I have shown for certain dissipative quantum maps that the traces of
(iterations of) the
propagator of the quantum mechanical 
density matrix agrees  to first order in an asymptotic expansion in
$\hbar$ with the traces of the classical Frobenius--Perron propagator of
the phase space density  if a small amount of dissipation is present. This
holds in spite of the fact 
that the corresponding spectra are very  different. I have 
tested the theory numerically for finite values of $\hbar$ for a dissipative
kicked top and have found good agreement 
in parameter regimes that ranged
from very weak to strong dissipation. The phase space structure turned out
to be important in the sense that higher quantum mechanical traces agree
with very high 
precision with the classical ones if the phase space is dominated by a point
attractor (strong dissipation),
whereas the precision is lost for higher traces in the case of an extended
strange attractor (weak dissipation). Sufficiently far away from
bifurcations the lowest traces always agree very
well with their classical counterpart.

{\it Acknowledgments:} I gratefully acknowledge fruitful discussions with
P.A.Braun, F.Haake, and J.Weber, and hospitality of the ICTP Trieste, where
part of 
this work was done.
Numerical computations were partly performed at the
John von Neumann--Institute for Computing in J\"ulich.
  
\begin{appendix}
\section{Classical maps and their stability matrices}\label{appA}
I give here the classical maps for the three components rotation, torsion
and dissipation as well as their stability matrices in phase space
coordinates. All maps will be 
written in the notation $(\mu,\phi)\longrightarrow (\nu,\psi)$, i.e.~$\mu$
and $\nu$ 
stand for the initial and final momentum, $\phi$ and $\psi$ for the initial
and final (azimuthal) coordinate. The latter is defined in the
interval from $-\pi$ to $\pi$. The stability matrices will be arranged as 
\begin{equation} \label{Mgen}
M=\left(
\begin{array}{cc}
\frac{\partial \psi}{\partial \phi}&\frac{\partial \nu}{\partial \phi}\\
\frac{\partial \psi}{\partial \mu}&\frac{\partial \nu}{\partial \mu}\\
\end{array}
\right)\,.
\end{equation}
{\em a. Rotation by an angle $\beta$ about $y$--axis}\\
The map reads
\begin{eqnarray}
\nu&=&\mu\cos\beta-\sqrt{1-\mu^2}\sin\beta\cos\phi\\
\psi&=&\Big(\arcsin(\sqrt{\frac{1-\mu^2}{1-\nu^2}}\sin\phi)\theta(x')+\\
&&(\sign(\phi)\pi-\arcsin(\sqrt{\frac{1-\mu^2}{1-\nu^2}}\sin\phi)
\theta(-x')\Big)\mod 2\pi\\
x'&=&\sqrt{1-\mu^2}\cos\phi\,\cos\beta+\mu\sin\beta\,,
\end{eqnarray}
where $x'$ is the $x$ component of the angular momentum after rotation,
$\theta(x)$ the Heaviside theta--function, and $\sign(x)$ denotes the sign
function. \\
The  stability matrix connected with this map is 
\begin{equation} \label{Mr}
M_r=\left(
\begin{array}{cc}
\sqrt{1-\mu^2}\left(\frac{\cos\phi}{\sqrt{1-\nu^2}\cos\psi}+\frac{\nu\sin\phi\tan\psi\sin\beta}{1-\nu^2}\right)&
\,\,\sqrt{1-\mu^2}\sin\phi\sin\beta\\
\frac{\nu\sin\psi(\sqrt{1-\mu^2}\cos\beta+\mu\cos\phi\sin\beta)}{\sqrt{1-\mu^2}(1-\nu^2)\cos\psi}-\frac{\mu\sin\phi}{\sqrt{(1-\nu^2)(1-\mu^2)}\cos\psi}
&\cos\beta+\frac{\mu\cos\phi\sin\beta}{\sqrt{1-\mu^2}}
\end{array}
\right)\,.
\end{equation}
{\em b. Torsion about $z$--axis}\\
Map and stability matrix are given by
\begin{eqnarray}
\nu&=&\mu\\
\psi&=&(\phi+k\mu)\mod 2\pi\\
M_t&=&\left(
\begin{array}{cc}
1&0\\
k
&1\\
\end{array}
\right)\label{Mt}\,.
\end{eqnarray}
{\em c. Dissipation}\\
The dissipation conserves the angle $\phi$, and the stability matrix is
diagonal:
\begin{eqnarray}
\nu&=&\frac{\mu-\tanh\tau}{1-\mu\tanh\tau}\\
\psi&=&\phi\\
M_d&=&\left(
\begin{array}{cc}
1&0\\
0&\frac{1-(\tanh\tau)^2}{(1-\mu\tanh\tau)^2}\\
\end{array}
\right)\label{Md}\,.
\end{eqnarray}

The total stability matrix for the succession rotation, torsion, dissipation
is given by $M=M_dM_tM_r$.
\end{appendix}

\end{document}